\documentclass[aps,prl,twocolumn,superscriptaddress,longbibliography]{revtex4-1}
\usepackage{amsmath}
\usepackage{hyperref}
\hypersetup{
	colorlinks=true,
	linkcolor=blue,
	filecolor=blue,
	citecolor = blue,      
	urlcolor=cyan,
}

\usepackage{graphicx}
\usepackage{dcolumn}
\usepackage{bm}

\usepackage[utf8]{inputenc}
\usepackage[T1]{fontenc}
\usepackage{mathptmx}
\usepackage{amsfonts}
\usepackage{amsmath}

\begin{document}


%


\title{Light confinement by local index tailoring in inhomogeneous dielectrics}
\author{I.~Kre\v{s}i\'{c}}
 \email{ivorkresic@gmail.com}
\affiliation{Institute for Theoretical Physics, Vienna University of Technology (TU Wien), Vienna, A-1040, Austria}
\author{K. G. Makris}
\affiliation{ITCP-Physics Department, University of Crete, Heraklion, 71003, Greece}
\affiliation{Institute of Electronic Structure and Lasers (IESL), Foundation for Research and Technology - Hellas, Heraklion, 71110, Greece}
\author{S. Rotter}
\affiliation{Institute for Theoretical Physics, Vienna University of Technology (TU Wien), Vienna, A-1040, Austria}

\date{\today}

\begin{abstract}
The engineering of light confinement is a topic with a long history in optics and with significant implications for the control of light-matter interaction. In inhomogeneous and disordered media, however, multiple scattering prevents the application of conventional approaches for the design of light fields with desired properties. This is because any local change to such a medium typically affects these fields in a non-local and complicated fashion. Here, we present a theoretical methodology for tailoring an inhomogeneous one-dimensional (1D) Hermitian dielectric index distribution, such that the intensity profile of an incoming light field can be controlled purely locally, i.e., with little or no influence on the field profile outside of a designated region of interest. Strongly increasing or decreasing the light's intensity at arbitrary positions inside the medium thereby becomes possible without, in fact, changing the external reflectance or transmittance of the medium. These local modifications of the medium can thus be made undetectable to unidirectional far field measurements. We apply our approach to locally control the confinement of light inside 1D materials with inhomogeneous continuous refractive index profiles and extend it to multilayer films as well as to chains of coupled micro-resonators. 
\end{abstract}

\maketitle


\section{Introduction}\label{intro}

The propagation of light through dielectric structures can nowadays be tailored to a remarkable degree \cite{joannopoulos_photonic_2008,molesky_inverse_2018,wiersma_disordered_2013,yu_engineered_2020}. In this respect, the understanding of the spatial localization of electromagnetic fields inside complex dielectric geometries has proven highly relevant for engineering light-matter interactions. Progress in this direction has already resulted in a broad range of designs for optical devices such as lasers, switches and memories \cite{notomi_manipulating_2010}. The key benefit of using dielectric geometries for trapping and manipulating light is the miniaturization, which in turn enhances light-matter interaction and enables integration of optical elements in a compact structure \cite{cheben_subwavelength_2018}.

A common approach of confining light is to rely on resonant modes, for which the field intensity is concentrated in a given region of space, and drops sharply outside of it. In photonic crystals, such a localization can, e.g., be achieved by placing a defect inside the periodic lattice~\cite{joannopoulos_photonic_2008}. Structures that are much easier to fabricate and that also feature strongly localized resonances are disordered media, whose potential for applications has been widely studied in recent years \cite{wiersma_disordered_2013,yu_engineered_2020,sebbah_random_2002,yamilov_highest-quality_2004,conti_dynamic_2008,sapienza_cavity_2010,noh_control_2011,mosk_controlling_2012,spasenovic_measuring_2012,vynck_photon_2012,riboli_engineering_2014,rotter_light_2017,riboli_tailoring_2017}. The multiple scattering of light, however, prevents a straightforward control of the resonances inside these media such that engineering their location and shape both in the spatial and in the spectral domain remains difficult. This is due to the fact that any local change of the refractive index of the disordered medium will have a considerable influence on the entire intensity distribution of a wave propagating through this medium (provided, of course, that the intensity is not negligible in the region where the refractive index is changed). This non-local dependence of the field's intensity on the entire scattering environment renders any attempt to optimize the field pattern a challenging and potentially impossible task.

In this article we demonstrate how this severe limitation, which constitutes a major hurdle for the design of complex media, can be conveniently overcome. We present an exact approach for tailoring the wave's intensity profile inside an inhomogeneous one-dimensional (1D) Hermitian dielectric, and apply it to engineer the confinement of light in disordered structures. Our methodology is based on a local, but judiciously calibrated modification of the medium's refractive index distribution by mapping the scattering fields of the given structure (reference medium) onto those of the desired structure (design medium). Specifically, these two scattering fields are proportional to each other with a space-dependent proportionality factor that we can choose to be strongly peaked or reduced in any desired region of space. The flexibility of our method allows us not only to select the exact position where such states are localized, but also to precisely engineer the shape of the corresponding intensity peaks by tailoring the shape of the local dielectric function. The spatial extent of the engineered intensity peaks is not limited to the scale of the optical wavelength, but can be made arbitrarily small (in principle). Notably, to achieve these effects, we do not need to work with non-Hermitian materials for which finite values of gain or loss would be necessary~\cite{yu_bohmian_2018,brandstotter_scattering-free_2019,makris_scattering-free_2020,tzortzakakis_shape-preserving_2020}, nor are any other exotic materials required, that would feature properties like non-reciprocal transmission characteristics or a vanishing/negative index of refraction. More specifically, the Hermitian media we consider here are fully described by real valued permittivity profiles.  

One specific consequence of the mapping we employ in our approach is that the reflection coefficients of the original (reference) and the modified (design) medium at the design wavenumber of the incident wave are equal (modulo $2\pi$ in the phase). Moreover, the transmission coefficients of these two media have equal moduli and can be engineered to have also equal phases. This makes the two interrelated media indistinguishable to unidirectional far field measurements for incoming light at the design wavevector. This aspect is highly interesting from the standpoint of concealing and mimicking the reflection and transmission signals of scattering objects. By establishing a connection between fields in an inhomogeneous reference medium and a design medium with equal unidirectional scattering coefficients, our optical design technique generalizes the interesting results of Refs.\ \cite{liu_direct_2017,king_designing_2018}, where the amplitude manipulation was limited to homogeneous reference media only, which considerably simplified the problem. 
\begin{figure}[t!]
\centering
\includegraphics[clip,width=1.0\linewidth]{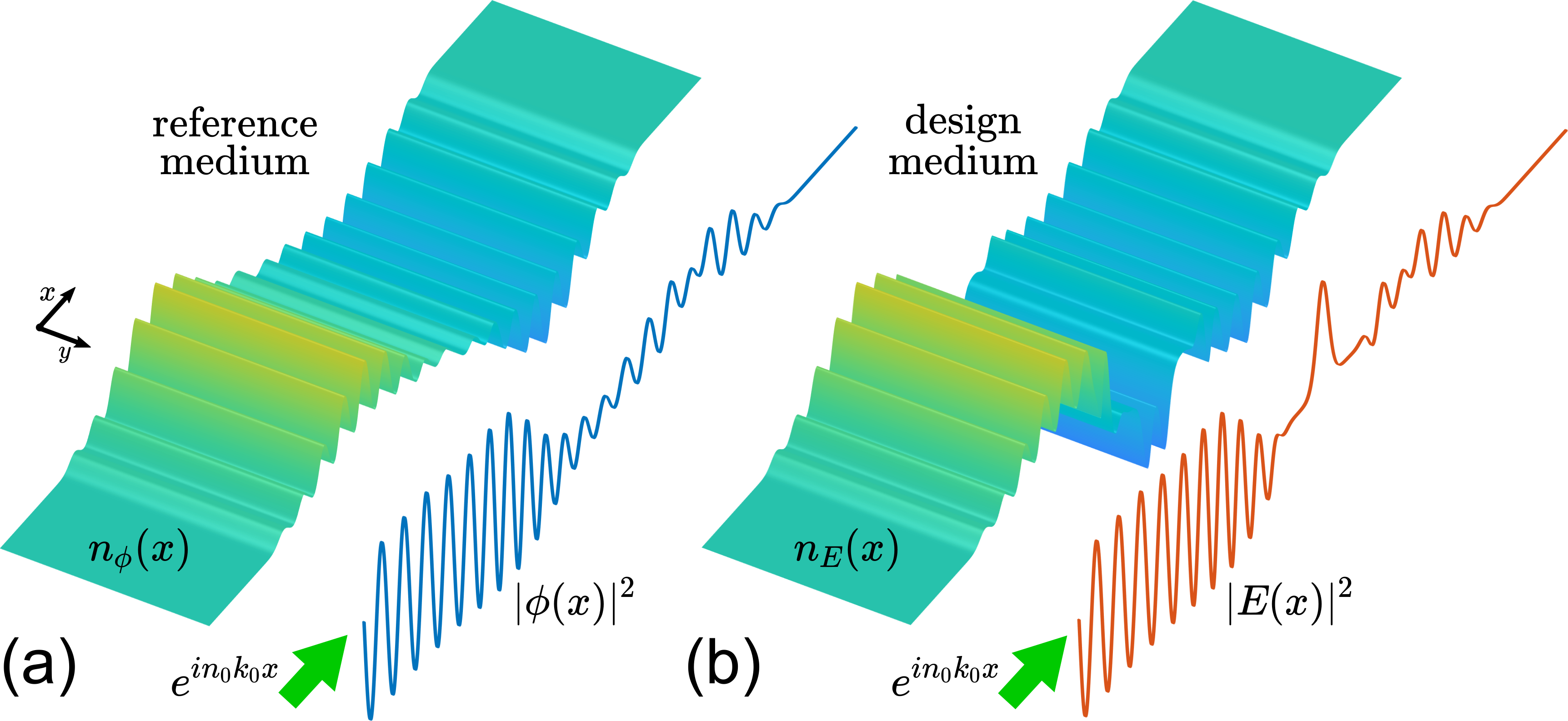}
\caption{Schematic depiction of our optical design principle. (a) A plane wave $e^{in_0k_0x}$ with wavenumber $k_0=2\pi/\lambda_0$ enters an inhomogeneous dielectric medium (left) with a refractive index $n_\phi(x)$ ($n_0$: refractive index of the homogeneous background medium, $\lambda_0$: wavelength of light in vacuum). The complex scattering processes inside the medium give rise to a highly modulated electric field intensity distribution $|\phi(x)|^2$ (right). (b) By locally modifying the refractive index $n_\phi(x)$ according to Eq. (\ref{eq:eps_mod}), we create a new medium $n_E(x)$ (left), which has the desired property of light confinement at the center, as depicted in the intensity distribution $|E(x)|^2$ (right), for the same input. Moreover, outside of the confinement region, the electric field $E(x)$ is indistinguishable in amplitude from the field $\phi(x)$ (compare to (a), right, and see Fig. \ref{Fig:2}a,b), and can also be made indistinguishable in phase from the field $\phi(x)$.}\label{Fig:1}
\end{figure} 
\section{Mapping between two inhomogeneous Hermitian media}\label{sec:two}
The problem we investigate is schematically depicted in Fig.~\ref{Fig:1}: we consider the propagation of light through linear, non-magnetic dielectrics that are inhomogeneous, but isotropic in the sense that the permittivity tensor can be replaced by a spatially-dependent scalar permittivity. For the case that this permittivity varies only along one direction (labeled as the $x$-coordinate in our case), but stays constant in the two orthogonal directions ($y,z$)—such as for layered media — the  Maxwell equations governing the light fields can be reduced to the much simpler Helmholtz equation in one dimension ($x$, see below). Specifically, the Helmholtz equation describes the orthogonal component of the time-harmonic electric field ($E_z$) propagating in $x$-direction at a fixed frequency $\omega_0=ck_0$, where $c$ is the vacuum speed of light, and $\lambda_0$ is the vacuum wavelength of light with $k_0=2\pi/\lambda_0$ being the corresponding wavenumber. Our methodology provides a systematic way to locally modify the dielectric function $\varepsilon_\phi(x)=n_\phi^2(x)$ of the reference medium in order to achieve a desired electric field intensity distribution inside a design medium with a dielectric function $\varepsilon_E(x)=n_E^2(x)$. 

When both media satisfy the same general material properties outlined at the beginning of the previous paragraph, the corresponding equations for the propagation of the electric field $\mathbf{\phi}(\mathbf{r})$ in the reference medium $\varepsilon_\phi(x)$, and of the electric field $\mathbf{E}(\mathbf{r})$ in the design medium $\varepsilon_E(x)$ read as follows \cite{joannopoulos_photonic_2008,yariv1984optical}: 
\begin{equation}\label{eq:Maxwell}
\begin{array}{c}
 \nabla\times\nabla\times\mathbf{\phi}(\mathbf{r})=k_0^2\varepsilon_\phi(x)\mathbf{\phi}(\mathbf{r}) ,\\[0.2cm]
 \nabla\times\nabla\times\mathbf{E}(\mathbf{r})=k_0^2\varepsilon_E(x)\mathbf{E}(\mathbf{r}).
\end{array}
\end{equation}
For waves incident in the $x$-direction, their polarized field component  $\mathbf{\phi}(\mathbf{r})=\phi(x)\mathbf{\hat{z}}$ and $\mathbf{E}(\mathbf{r})=E(x)\mathbf{\hat{z}}$ is described by the following 1D Helmholtz equations for the two media:
\begin{equation}\label{eq:Helm}
\begin{array}{c}
 \left[\frac{d^2}{dx^2}+k_0^2\,\varepsilon_\phi(x)\right]\phi(x)=0  ,\\[0.2cm]
\left[\frac{d^2}{dx^2}+k_0^2\,\varepsilon_E(x)\right]E(x)=0.
\end{array}
\end{equation}

As depicted in Fig.~\ref{Fig:1}, our goal is to associate the reference medium with $\varepsilon_\phi(x)\in \mathbb{R}$ to a design medium $\varepsilon_E(x)\in \mathbb{R}$, such that the corresponding scattering solutions are related as:
\begin{equation}\label{eq:rel}
E(x) = R(x)\phi(x),
\end{equation}
where the space-dependent factor $R(x) = A(x)e^{i\theta(x)}$ is complex-valued and $A(x)$, $\theta(x)$ are the real-valued functions of amplitude and phase, respectively. It can be shown (see Supporting information-SI), that if both media are Hermitian, the modified dielectric function $\varepsilon_E(x)$ is related to the reference dielectric function $\varepsilon_\phi(x)$ by:
\begin{equation}\label{eq:eps_mod}
\varepsilon_E(x) = \varepsilon_\phi(x)+\Delta\varepsilon(x),
\end{equation}
where
\begin{equation}
\Delta \varepsilon(x) = \frac{1}{k_0^2}\left[\left(\frac{d\theta}{dx}\right)^2+2\eta_I\frac{d\theta}{dx} -\frac{1}{A}\frac{d^2 A}{dx^2} -\frac{2\eta_R}{A}\frac{dA}{dx} \right],
\end{equation}
with $\frac{d}{dx}\ln\phi=\eta_R+i\eta_I$, and $\theta(x)$ satisfying the equation:
\begin{equation}\label{eq:theta}
\frac{d^2\theta}{dx^2}+2\left(\frac{1}{A}\frac{dA}{dx}+\eta_R\right)\frac{d\theta}{dx}+\frac{2\eta_I}{A}\frac{dA}{dx}=0.
\end{equation}
 \begin{figure}[t!]
\centering
\includegraphics[clip,width=1\linewidth]{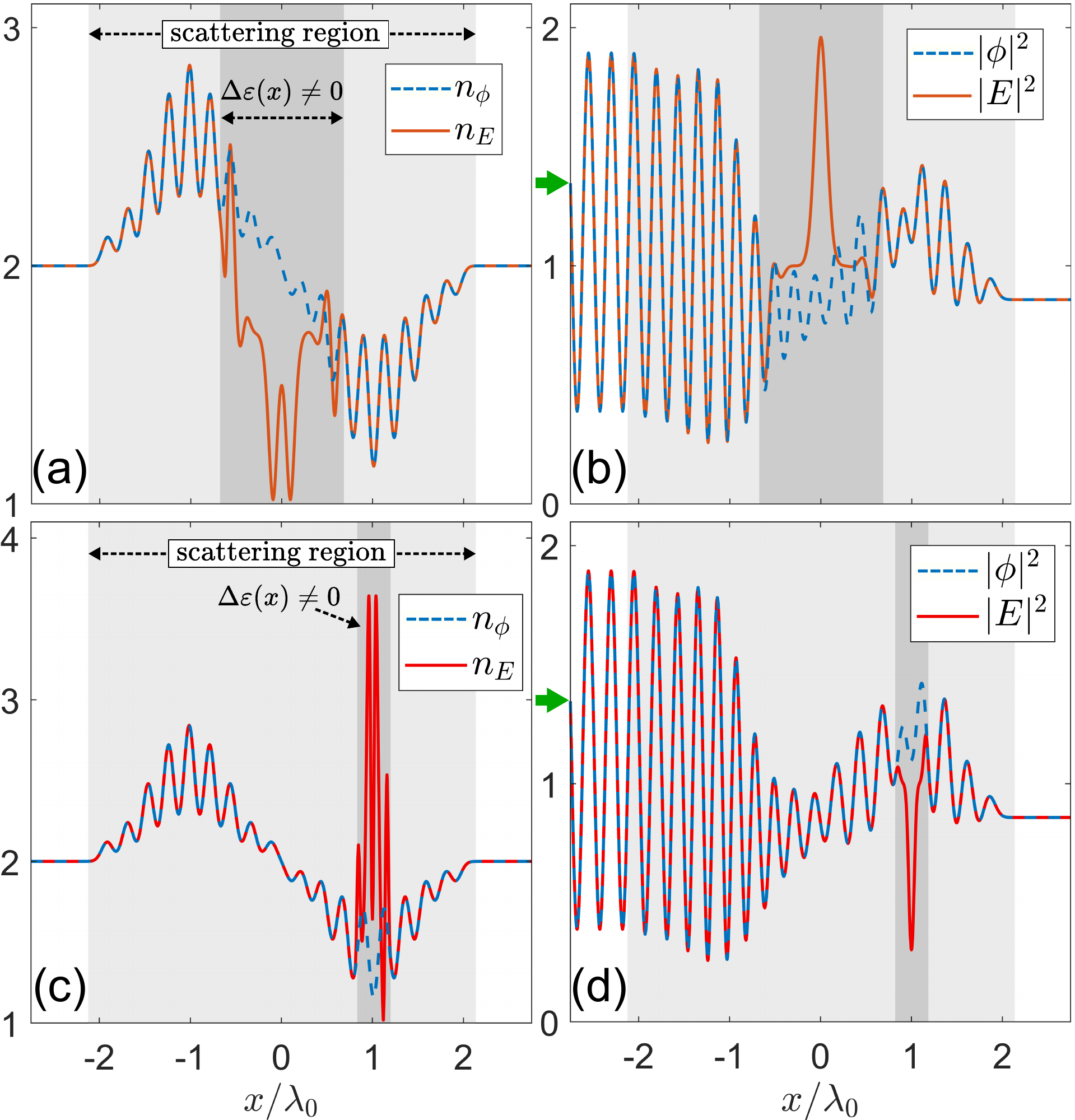}
\caption{Local intensity control inside an inhomogeneous scattering medium. (a) Refractive index distribution of the reference medium $n_\phi(x)$ (blue dashed) and of the designed medium $n_E(x)$ (orange solid), leading to light confinement of a Gaussian intensity profile centered around $x=x_c=0$. (b) Normalized electric field intensity for the reference medium, $|\phi(x)|^2$ (blue dashed) and for the designed medium, $|E(x)|^2$  (orange solid), corresponding to the refractive index distributions depicted in (a). (c) Refractive index distribution of the reference medium $n_\phi(x)$ (blue dashed) and the designed medium $n_E(x)$ (red solid), for which the intensity of light is decreased at $x=x_{c}=\lambda_0$, as shown in (d). (d) Normalized electric field intensities for the reference medium, $|\phi(x)|^2$ (blue dashed) and for the designed medium, $|E(x)|^2$ (red solid) with refractive index profiles depicted in (c). In all plots the areas shaded in light gray color mark the scattering region, while the dark shaded parts indicate the region where also $\Delta \varepsilon(x)\neq 0$. The green arrows in (b) and (d) denote the incident waves' propagation direction. The refractive index profile of the reference medium was constructed as a superposition of $N=18$ Gaussians with equal widths and spacing, but with varying amplitudes, such that $n_\phi(x)=n_0+f(x)$, where $f(x)=-f(-x)$. $A(x)$ is given by Eq.\ (\ref{eq:Amp}), while the parameters for the plots (a) and (b) are: $x_{c}=0$, $\sigma=0.55\lambda_0$, $\alpha=0.4$, $\sigma_{c}=0.07\lambda_0$, where $\lambda_0$ is the vacuum wavelength. The parameters for the plots (c) and (d) are: $\alpha=-0.45$, $\sigma_{c}=0.0265\lambda_0$, $x_{c}=\lambda_0$, $\sigma=0.15\lambda_0$.}\label{Fig:2}
\end{figure}

In this article, we solve Eq.\ (\ref{eq:theta}) numerically as a differential equation for the relative phase $\theta(x)$ inside the scattering medium for a given amplitude function $A(x)$, which we call the design function. The equation can also be reformulated as an equation for $A(x)$ at a given $\theta(x)$ (see Ref.\ \cite{king_designing_2018}), however we do not follow this approach here. An equivalent formulation of the problem, which does not require the numerical solution of Eq.\ (\ref{eq:theta}), is given in Section 3 of the SI. However, such a formulation makes the relationship between the solutions in terms of $n_\phi(x)$ and $n_E(x)$ less transparent, so it was not used for the basic results of this article (see Section \ref{sec:four} and the SI). Equation (\ref{eq:theta}) stems from the fact that the designed medium is Hermitian, and a related equation in Ref.\ \cite{king_designing_2018} was aptly named the ``energy conservation condition''. The coefficients $\eta_R(x),\:\eta_I(x)$ are calculated from the numerical solution of the Helmholtz equation for the reference medium. Here we consider functions $A(x)$, which approach the value one outside the modified region (see below), which means that outside of this region $|\phi(x)|=|E(x)|$.

Since we demand $\varepsilon_\phi(x\to\pm\infty)=\varepsilon_E(x\to\pm\infty)=n_0^2$ (where $n_0$ is the homogeneous background refractive index), we numerically solve Eq.\ (\ref{eq:theta}) with imposed Neumann boundary conditions $\frac{d\theta(x)}{dx}=0$ at the boundaries of our system. It can be shown that for these boundary conditions, $\theta(x)$ is always constant and $\varepsilon_\phi(x)=\varepsilon_E(x)$ in the region where $A(x)= 1$ (see SI).

For the wavenumber $k=k_0$, the reflection and transmission coefficients of right-propagating incident waves for the two media are intimately connected with each other through Eq.\ (\ref{eq:theta}) (see SI):
\begin{equation}\label{eq:ref_tran}
r_E(k_0)=r_\phi(k_0)\:\:(\mbox{mod $2\pi$ in phase}),\:\: t_E(k_0)=e^{i\delta_t(k_0)}\,t_\phi(k_0),
\end{equation}
where $\delta_t(k_0)$ is the relative phase shift acquired in the design medium with respect to the reference medium. Additionally, it follows from the Hermiticity of the scattering problem \cite{albert_messiah_quantum_1958} that the scattering coefficients of the two media for left-propagating waves are related as:
$|r_E'(k_0)|=|r_E(k_0)|=|r_\phi(k_0)|=|r_\phi'(k_0)|$ and $|t_E'(k_0)|=|t_E(k_0)|=|t_\phi(k_0)|=|t_\phi'(k_0)|$. We note that, although in this article we consider solutions $\phi(x)$ with right-propagating incident waves, the same theory could also be used for left-propagating incident waves. Also in this case the design condition Eq.\ (\ref{eq:rel}), leads to Eq.\ (\ref{eq:theta}) for Hermitian media, and thus determines the relation between the asymptotic properties of the family of dielectric functions related by Eq.\ (\ref{eq:eps_mod}). 

Most importantly, for a judiciously chosen design function $A(x)$, the unidirectional shift $\delta_t(k_0)$ can be completely eliminated (see below). In other words, our strategy should also allow us to choose the design medium $n_E(x)$ in such a way that 
it cannot be distinguished from the reference medium $n_\phi(x)$ by unidirectional far-field measurements, i.e., $E(x\to\pm\infty)=\phi(x\to\pm\infty)$ for either a right- or left-propagating incident plane wave with a wavenumber $k_0$.

We now apply the above approach to increase or reduce the intensity of light inside a 1D Hermitian dielectric by locally modifying its refractive index profile $n_\phi(x)$. As our first example, we consider the following form for the design function $A(x)$:
\begin{equation}\label{eq:Amp}
A(x)=1-\left( 1-\frac{1}{|\phi(x)|} \right)e^{-(x-x_{c})^8/\sigma^8}+\frac{\alpha}{|\phi(x)|}e^{-(x-x_{c})^2/2\sigma_{c}^2}.
\end{equation}
The first two terms of the above expression affect the field profile $|E(x)|$, since they produce a flat (super-Gaussian) region of height equal to one and width $\sigma$ centered at $x=x_{c}$, while the last term adds to this background a Gaussian function with amplitude $\alpha$ and width $\sigma_{c}$ at the same center position. To obtain a Gaussian shaped confinement, it is necessary to add the super-Gaussian term, since otherwise the confinement region would have a background varying as $|\phi(x)|$. Depending on the sign of $\alpha$, the solution is increasing or reducing the light's intensity at $x=x_{c}$, as shown in the two examples of  Fig.\ \ref{Fig:2}. 
\begin{figure}[t!]
\centering
\includegraphics[clip,width=1\linewidth]{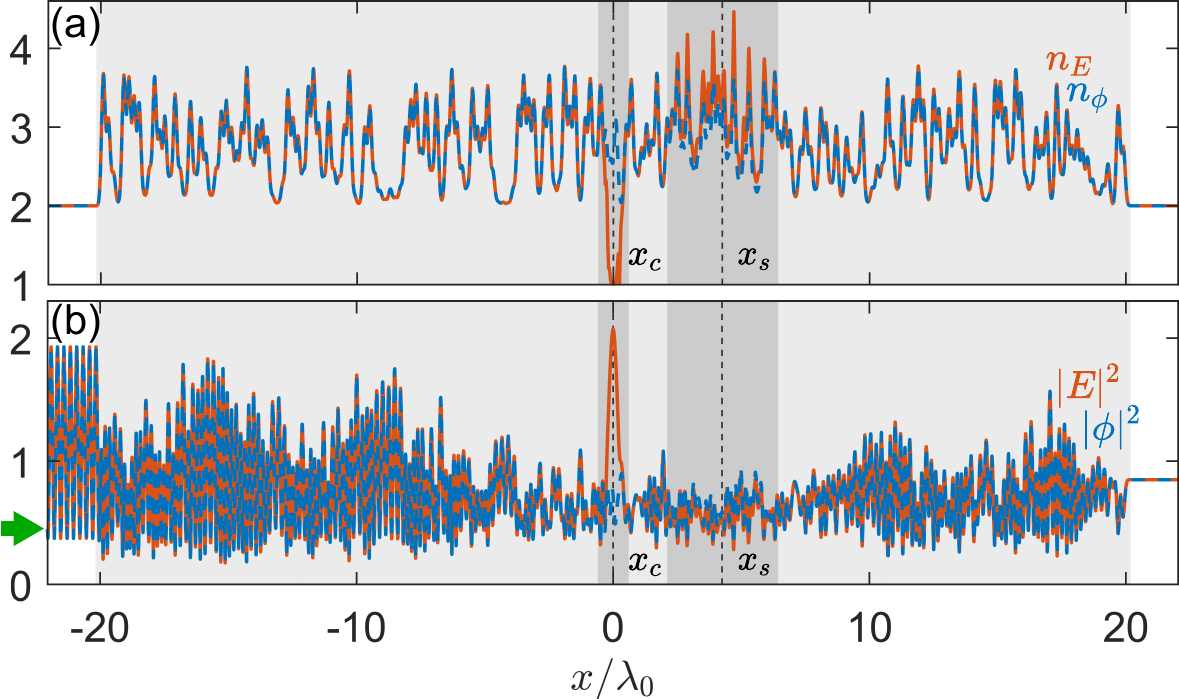}
\caption{Confining light at the middle of a disordered structure by modifying the reference medium based on a design function $A(x)+\delta a(x)$ [see Eqs.\ (\ref{eq:Amp}) and (\ref{eq:Ampdelta})], that leaves the complex transmission coefficients for a right-propagating incident wave at $k=k_0$ unchanged (see Fig.\ \ref{Fig:4}). (a) Refractive index profiles of the reference $n_\phi(x)$ (blue dashed) and of the design medium $n_E(x)$ (orange solid). (b) Electric field intensities (normalized to input) of the reference $|\phi(x)|^2$ (blue dashed) and in the design medium $|E(x)|^2$ (orange solid). As in Fig.\ \ref{Fig:2}, the light gray areas mark the scattering region, while the dark gray areas mark the regions where also $\Delta\varepsilon(x)\neq 0$. The green arrow in (b) denotes the propagation direction of the incident wave. The dashed vertical lines mark $x_{c}$ and $x_{s}$ (see text). The refractive index of the reference medium is a superposition of $N=200$ Gaussians with randomly varying amplitudes. The relevant parameters of the designed medium are: $\alpha=0.44$, $\sigma_{c}=0.15\lambda_0$, $x_{c}=0$, $\sigma=0.225\lambda_0$, $\beta = 0.0325041$, $x_{s}=4.25\lambda_0$, $\sigma_{s}=1.75\lambda_0$.}\label{Fig:3}
\end{figure} 
In particular, Fig.\ \ref{Fig:2} demonstrates the principle of local intensity engineering based on a refractive index modification  around $x=x_{c}$. In the first case, the refractive index in the dark grey region reduces below the background value $n_0=2$, forming a structure that resembles a well. The interference of the waves reflecting at the edges of the well then creates a region with a peaked electric field intensity. The well is shaped precisely such that the intensity peak is a Gaussian, while the reflection and transmission properties of this well structure result in an intensity distribution outside of the well that is identical to the one of the reference medium [see Eq.\ (\ref{eq:ref_tran})]. In the opposite case of intensity suppression at $x_{c}=\lambda_0$, the refractive index rises above the background value, creating a barrier. The field amplitude inside the barrier is reduced and forms a trough with a Gaussian shape. The barrier has such a shape that the transmitted and reflected waves have the same intensity distribution. For all the scattering problems of Fig.\ \ref{Fig:2} (and also in the rest of the article), the refractive index distribution $n_E(x)$ in the regions where $A(x)=1$ (so $\Delta\varepsilon(x)=0$) is equal to that of the reference medium $n_\phi(x)$, as is expected (see above). 
\begin{figure}[t!]
\centering
\includegraphics[clip,width=1.0\linewidth]{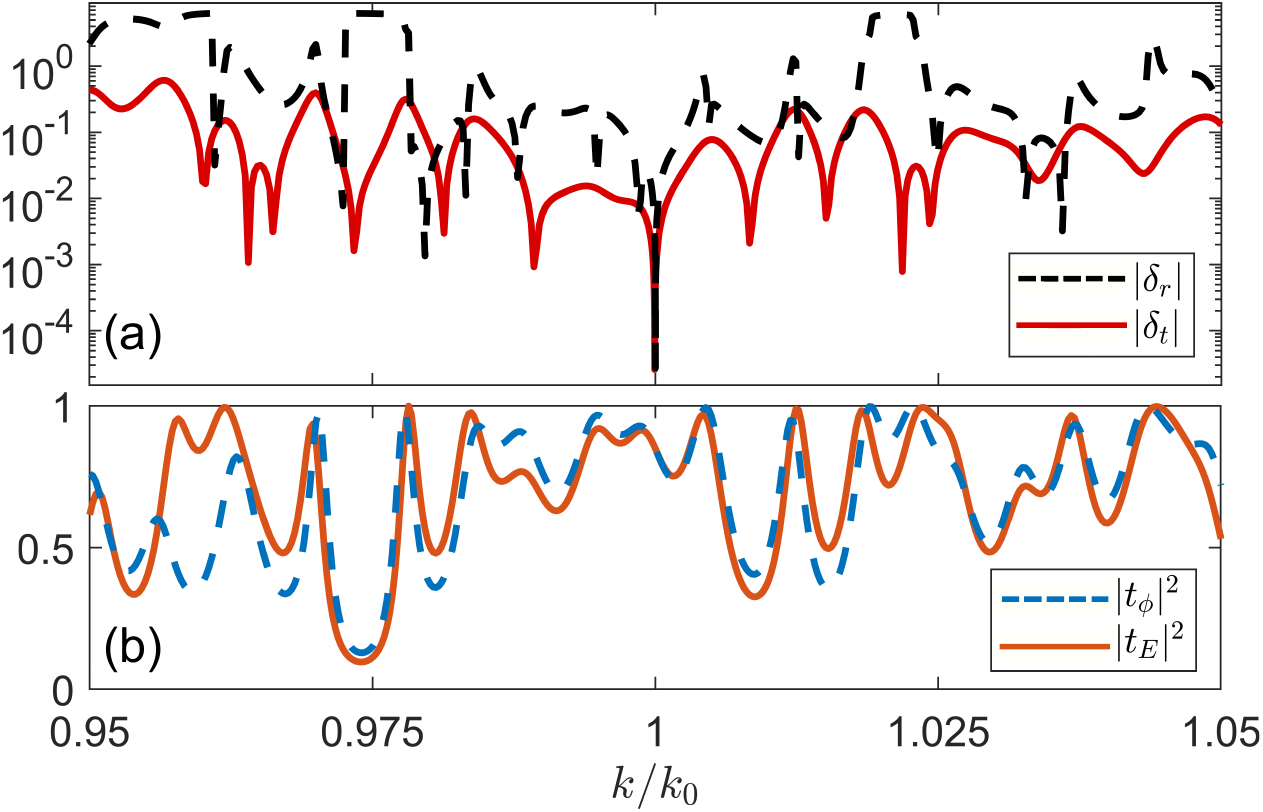}
\caption{Sensitivity of the scattering characteristics to variations of the wavenumber $k$, for the reference and the designed refractive index distributions shown in Fig.\ \ref{Fig:3}a. (a) Modulus of the difference of reflection phases ($|\delta_r(k)|=|\arg[r_E(k)]-\arg [r_\phi(k)]|$, black dashed) and transmission phases ($|\delta_t(k)|=|\arg[t_E(k)]-\arg [t_\phi(k)]|$, red solid) for the two media (in radians). (b) Modulus squared of the transmission coefficient for the reference (blue dashed) and the design medium (orange solid). The parameters for the design medium are given in the caption of Fig. \ref{Fig:3}.
}\label{Fig:4}
\end{figure} 

With our methodology we can thus, in principle, modify any reference refractive index distribution $n_\phi(x)$ and its associate scattering wave solution $\phi(x)$ with finite and continuous $\eta_R(x)$ and $\eta_I(x)$ (where $\frac{d}{dx}\ln\phi=\eta_R+i\eta_I$). However, for a Hermitian $n_\phi(x)$, the modified medium $n_E(x)$ will not be dielectric and Hermitian in general, since this depends on the specific parameters of the modified electric field intensity distribution. For instance, if one wants to confine light inside a Hermitian dielectric, then increasing the amplitude of the confined light beyond a certain value will result in the refractive index $n_E(x)$ falling below unity and/or having complex values at some region of space. In this article we consider only parameters for which both $n_\phi(x)$ and $n_E(x)$ correspond to Hermitian dielectric media.

\section{Unidirectionally indistinguishable disordered media}\label{sec:three}
The most striking application of our novel approach is related to the indistinguishability of disordered media. More specifically, we consider a 1D disordered medium that was constructed by a random superposition of many Gaussian functions, as the one depicted in Fig.\ \ref{Fig:3}. Whereas the media in the previous section were indistinguishable for far-field measurements of the \textit{intensity} (at $k=k_0$), we will now also remove any effect on the transmission \textit{phase} induced by the modification of the medium ($\delta_t(k_0)=0$) in the case of light confinement (see Fig.\ \ref{Fig:3}). To achieve this, we add to the design function $A(x)$ of Eq.\ (\ref{eq:Amp}) a part $\delta a(x)$ of the form:  
\begin{equation}\label{eq:Ampdelta}
\delta a(x)=-\frac{\beta}{|\phi(x)|} e^{-(x-x_{s})^8/\sigma_{s}^8},
\end{equation}
where $\beta>0$. Adding this contribution $\delta a(x)$ to the design function $A(x)$ creates a barrier in the refractive index distribution at position $x=x_{s}$. We have numerically tuned the height $\beta$ of the barrier such that it compensates for the phase shift caused by the wave's confinement.

\begin{figure}[t!]
\centering
\includegraphics[clip,width=1.0\linewidth]{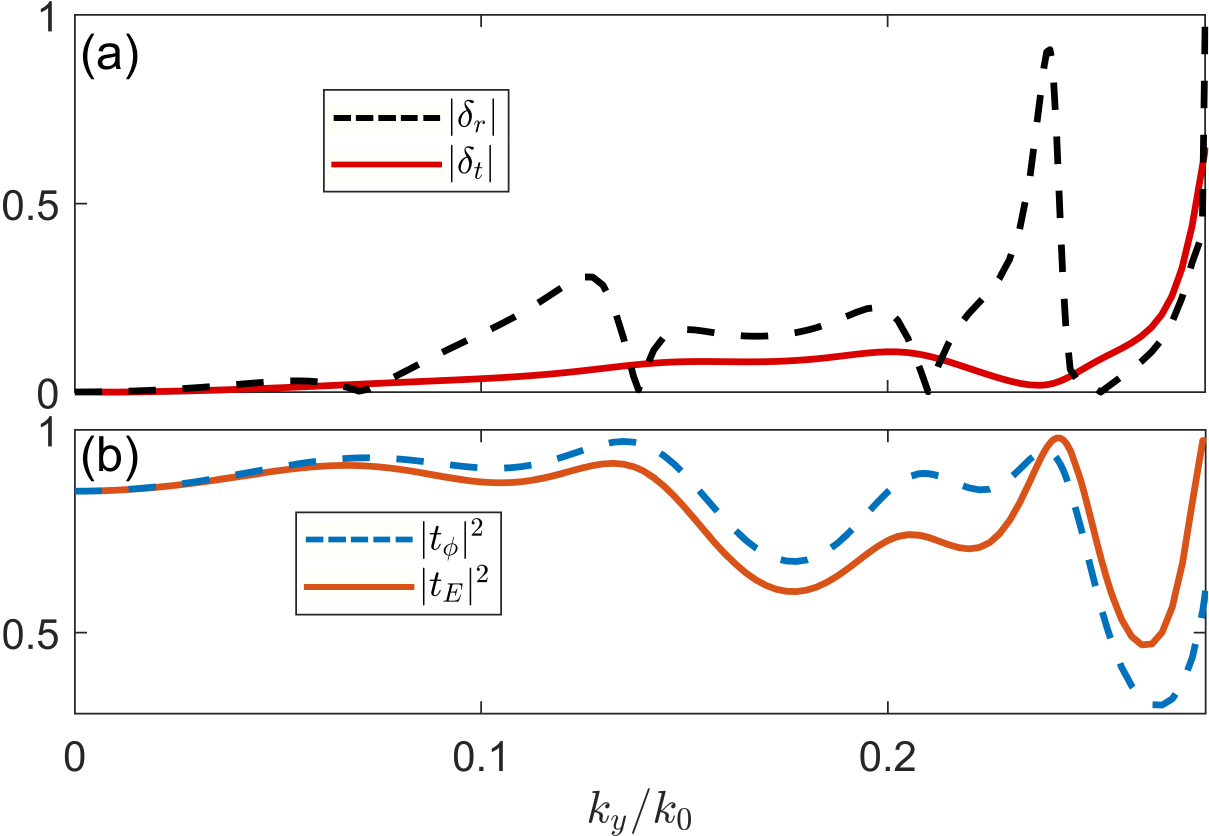}
\caption{Sensitivity of the scattering characteristics to variations in the incident beam's angle (quantified by the $y$-component of the input wavevector, $k_y$), for the reference and the designed refractive index distributions shown in Fig.\ \ref{Fig:3}a. (a) Modulus of the difference of reflection phases ($|\delta_r(k)|=|\arg[r_E(k)]-\arg [r_\phi(k)]|$, black dashed) and transmission phases ($|\delta_t(k)|=|\arg[t_E(k)]-\arg [t_\phi(k)]|$, red solid) for the two media (in radians). (b) Modulus squared of the transmission coefficient for the reference (blue dashed) and the design medium (orange solid). The parameters for the design medium are given in the caption of Fig. \ref{Fig:3}.
}\label{Fig:5}
\end{figure} 

Figure \ref{Fig:3}a clearly shows how the reference medium is modified in this case. As explained above, the confinement of light is achieved by the constructive interference induced by the dip in the refractive index around $x=x_{c}$. The shape of the dip is precisely tailored to produce the desired profile of the confined field, while maintaining the same amplitude as in the reference medium outside of the confinement region. The additional part $\delta a(x)$ now modifies the medium only slightly, in a region behind the confinement (for $x_{c}<x_{s}$). Figure \ref{Fig:3}b displays a perfect correlation between $|E(x)|^2$ and $|\phi(x)|^2$, apart from the confinement region centered at $x_{c}$ and inside the phase shifting region around $x_{s}$, where a small deviation between these two quantities is observable. Moreover, around $x=x_c$, the refractive index $n_E(x)$ is tailored to reach the value of 1. This part of the modified medium is free space (air), which can be accessed by small emitters such as quantum dots or atoms, enabling them to couple to the engineered electric field intensity $|E(x_c)|$ (see, e.g., Ref.\ \cite{chang_colloquium_2018}).

The above methodology for tailoring the intensity profile is a delicate wave interference effect and the explicit analytical relation between the reference and the design medium, Eq.\ (\ref{eq:eps_mod}), holds for a specific design wavenumber $k_0$. To test the spectral robustness of our procedure, we examine how the solutions of the Helmholtz equation in the reference and in the design medium change when the wavenumber $k$ of the incident wave is detuned away from the design wavenumber $k_0$.

In particular, the absolute value of the difference between reflection phases ($\delta_r(k)=\arg[r_E(k)]-\arg[r_\phi(k)]$) and transmission phases ($\delta_t(k)=\arg[t_E(k)]-\arg[t_\phi(k)]$) is plotted in Fig.~\ref{Fig:4}a for the case of $A(x)+\delta a(x)$ of Fig.\ \ref{Fig:3}. The quantities $|\delta_r(k)|$ and $|\delta_t(k)|$ reach the minimum values of $2.8\times10^{-5}$ and $2.6\times10^{-5}$ at $k=k_0$, respectively, and rapidly increase when moving away from $k_0$. The sharpness of the troughs in $|\delta_r(k_0)|$ and $|\delta_t(k_0)|$ are a signature of the fact that the effects leading to the confinement of light are interferometric in origin. Moving even further away from $k=k_0$, we observe discrete values of $k$ for which $|\delta_r(k)|$ and $|\delta_t(k)|$ drop or rise again sharply, indicating interesting resonance effects. 
\begin{figure}[t!]
\centering
\includegraphics[clip,width=1\linewidth]{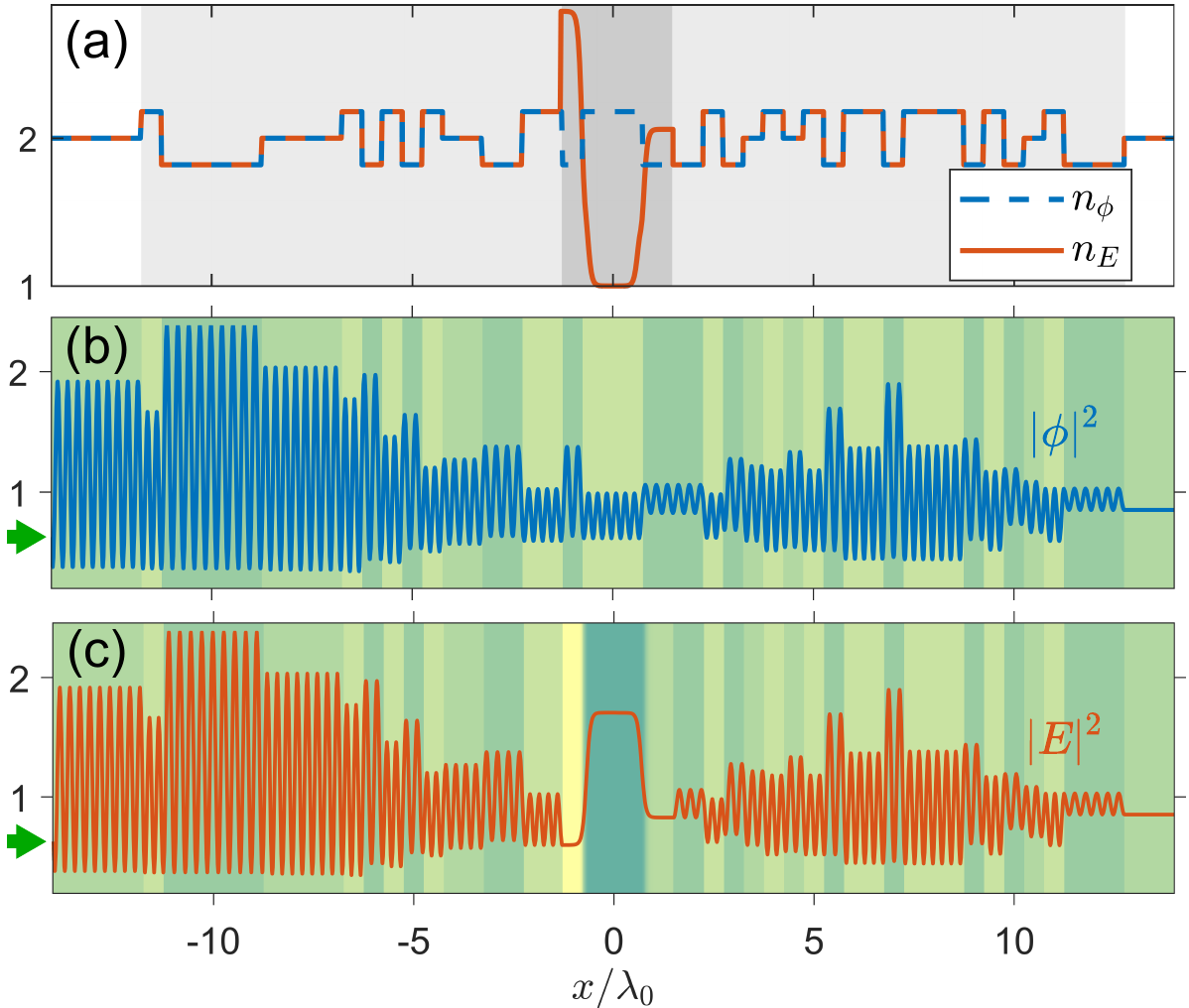}
\caption{Demonstrating light confinement by modifying a 1D stack of dielectric slabs. We directly apply our method, by choosing an amplitude design function $A(x)$, as given in Eq.\ (\ref{eq:A_discrete}). (a) Refractive index distributions of the reference medium, $n_\phi(x)$ (blue dashed) and of the design medium, $n_E(x)$ (orange solid). The refractive index $n_\phi(x)$ describes a collection of $N=50$ slabs of width $0.5\lambda_0$, alternating randomly between the values of 1.82, 2 and 2.18. The area shaded in light gray denotes the scattering region, while the dark shaded part is the region where the media differ, i.e., $\Delta\varepsilon(x)\neq 0$. Electric field intensity (normalized to input) for the solution (b) in the reference medium, $|\phi(x)|^2$ (blue) and (c) in the design medium, $|E(x)|^2$ (orange). The (false colored) density plots in the backgrounds of (b) and (c) are schematic depictions of $n_\phi(x)$ and $n_E(x)$, respectively. The green arrows in (b) and (c) indicate the propagation direction of the incident wave. The related parameters of the designed medium are: $x_L=-2.583\lambda_0$, $x_R=2.997\lambda_0$, $\alpha =0.533$, $x_{c}=0$, $L_f=0.7\lambda_0$ and $\sigma_{c}=0.05\lambda_0$.}\label{Fig:6}
\end{figure}
\begin{figure}[t!]
\centering
\includegraphics[clip,width=1\linewidth]{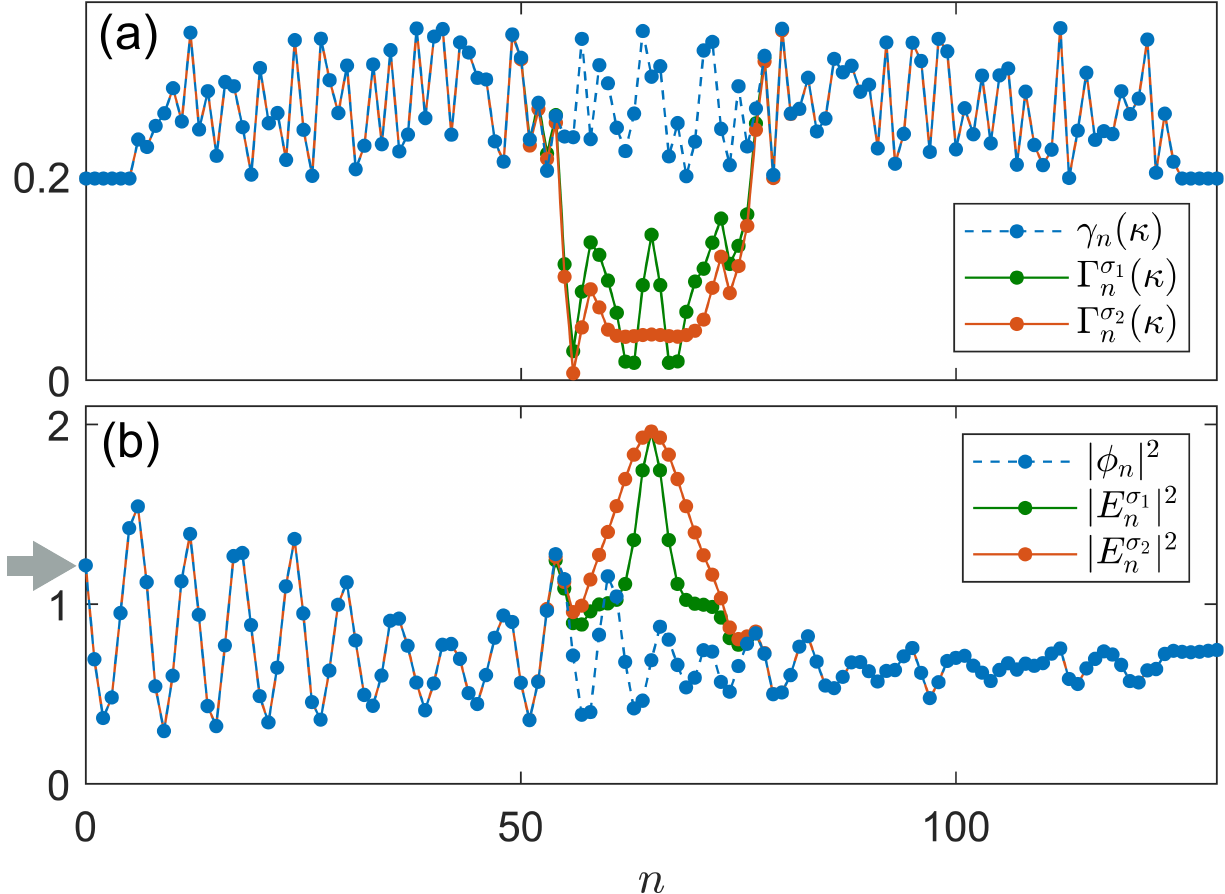}
\caption{Light confinement at the center of a chain of $N=130$ microresonators with a disordered resonance frequency distribution. (a)~Distribution of the microresonator resonance frequencies for the reference medium, $\gamma_n$ (blue dots) and for two design media with a width $\sigma_{c} =1.5$ ($\Gamma_n^{\sigma_1}$, green dots) and $\sigma_{c}=4$ ($\Gamma_n^{\sigma_2}$, orange dots), respectively. (b) Field intensity (normalized to input) inside the resonator chain for the reference medium, $|\phi_n|^2$ (blue dots) and the designed media for the two cases above, $|E_n^{\sigma_1}|^2$ (green dots) and $|E_n^{\sigma_2}|^2$ (orange dots). The amplitude distribution $A_n$ is given by $A_n=1-\left( 1-\frac{1}{|\phi_n|} \right)e^{-(n-n_c)^8/\sigma^8}+\frac{\alpha}{|\phi_n|}e^{-(n-n_c)^2/2\sigma_{c}^2}$. 
The gray arrow marks the propagation direction of the incident wave. The other relevant parameters are: $\omega=2\kappa$, $\alpha=0.4$, $n_c=65$, $\sigma=7.1$, $k_0=4\pi$, $\Delta x =0.025$, $\gamma_N=0.1974\kappa$.}\label{Fig:7}
\end{figure}
Moreover, the amplitudes of the transmission coefficients of the reference medium $t_\phi$ and of the design medium $t_E$, are also shown in Fig.\ \ref{Fig:4}b. As we can see, they depend less sharply on the variation of $k$, as compared to the changes on the transmission phase. This is a consequence of the particular choice of the refractive index profile and of the fact that the media with higher index contrasts are expected to be more sensitive to variations of $k$. Likewise, concerning the sensitivity of the media to design imperfections, we have found that several factors, such as the the refractive index contrast, the length scale of variations in the refractive index, the degree of light confinement and the strength of disorder in the system, all play a role in the robustness of our design. We do note, however, that, since the methodology we apply to design $n_E(x)$ is fairly general (including several free parameters), it is expected that the unidirectional indistinguishability of $n_\phi(x)$ and $n_E(x)$ can become broadband for an optimized choice of the design parameters.

To complement the above frequency scans, we have also performed scans of the input tilt angle, parametrized by the transverse wavevector component $k_y$, at the design frequency $\omega_0=ck_0$. By inserting the fields $\mathbf{\phi'}(\mathbf{r})=\phi'(x)e^{in_0k_yy}\mathbf{\hat{z}}$ and $\mathbf{E'}(\mathbf{r})=E'(x)e^{in_0k_yy}\mathbf{\hat{z}}$ into the Eqs.~(\ref{eq:Maxwell}), we get the equations for propagation of a tilted beam through media with dielectric functions $\varepsilon_\phi(x)$ and $\varepsilon_E(x)$, respectively:
\begin{equation}\label{eq:Helmky}
\begin{array}{c}
 \left[\frac{d^2}{dx^2}+k_0^2\,\varepsilon_\phi(x)-n_0^2k_y^2\right]\phi'(x)=0  ,\\[0.2cm]
\left[\frac{d^2}{dx^2}+k_0^2\,\varepsilon_E(x)-n_0^2k_y^2\right]E'(x)=0.
\end{array}
\end{equation}
\newline
The incoming field is now of the form $e^{in_0(k_xx+k_yy)}$, with $k_x$, $k_y$ satisfying the dispersion relation $k_0^2=k_x^2+k_y^2$. The refractive index distributions of the two media are plotted in Fig.~\ref{Fig:3}a.

The scans are presented in Fig.\ \ref{Fig:5}. Both the reflected ($\delta_r$) and transmitted ($\delta_t$) phase difference moduli stay below the value of 0.1 rad for $k_y$ values up to $0.09k_0$, corresponding to input angles of $|\alpha|=|\arctan(k_y/k_0)|=5.14^\circ$. Moreover, as the transmission coefficients $t_\phi$ and $t_E$ stay similar in both phase and amplitude (see Fig. \ref{Fig:5}b) for $k_y$ values up to $0.15k_0$, the initial and design medium are approximately indistinguishable to transmission measurements at input angles in the relatively broad range of $[-8.5^\circ,8.5^\circ]$ with respect to the normal. As the $k_x$ component of the input wavevector changes by $\approx 0.04k_0$ in the whole scan range of Fig. \ref{Fig:5}, the robustness of the transmission and reflection properties to changes of the incident beam's tilt is thus similar to the corresponding changes of the incident beam's frequency. 
\newline

\section{Potential experimental implementations}\label{sec:four}

In this section we study how our theoretical and numerical results could be in principle realized in an experiment. Certainly a challenging aspect is the fact that the assumed continuous refractive index profiles might be difficult to implement in practical photonic setups, due to their highly oscillatory $x$-dependence. We will thus investigate here, how to implement our theory in more realistic experimental settings. 

The first system we consider is a multilayer medium consisting of a 1D stack of dielectric slabs, which is a physical system that is frequently encountered in photonics \cite{joannopoulos_photonic_2008}. We have found that Eq.\ (\ref{eq:eps_mod}) can be applied to discontinuous $n_\phi(x)$ distributions. In particular, for certain forms of the amplitude design function $A(x)$, the resulting continuous refractive index distribution $n_E(x)$ can be made to vary without fast oscillations on a scale of the typical manufacturing resolution of photonic crystals. Implementing such $n_E(x)$-distributions in an experimental setting would be a fascinating prospect for photonic design. To illustrate such a potential implementation, we provide in the following one specific example of such an amplitude function $A(x)$: 
\begin{widetext}
\begin{equation}\label{eq:A_discrete}
\begin{array}{c}
    A(x) = \begin{cases}
             |\phi(x)|^{-1}\times\left\{|\phi(x_L)|+ \alpha\left([1+e^{(x_{c}-x-L_f)/\sigma_{c}}]^{-1}+[1+e^{(x-x_{c}-L_f)/\sigma_{c}}]^{-1}-1\right) \right\}         , & x_L\leq x< 0\\
             
             |\phi(x)|^{-1}\times\left\{|\phi(x_R)|+ (\alpha+|\phi(x_L)|-|\phi(x_R)|)\left([1+e^{(x_{c}-x-L_f)/\sigma_{c}}]^{-1}+[1+e^{(x-x_{c}-L_f)/\sigma_{c}}]^{-1}-1\right) \right\}        , & 0\leq x\leq x_R
               \end{cases}
\end{array}
\end{equation}
\end{widetext}
where $x_L$ and $x_R$ are the left and right boundaries of the region where $\Delta\varepsilon(x)\neq 0$, and $A(x<x_L)=A(x>x_R)=1$. We have found that by choosing the $x_{L,R}$ boundaries to be at the local minima of $|\phi(x)|^2$, the distribution $n_E(x)$ can be made to vary smoothly inside this region (see the SI for a discussion on designing smooth $n_E(x)$ functions). The results shown in Fig.\ \ref{Fig:6} demonstrate an example of using Eq.\ (\ref{eq:A_discrete}) to create light confinement by modifying a discontinuous distribution $n_\phi(x)$, describing a stack of randomly alternating dielectric slabs. In this example, the resulting intensity distribution $|E(x)|^2$ has a shape described by Eq. (\ref{eq:A_discrete}), giving light confinement at $x=x_c=0$ (see also Eq. (16) of the SI). As in Fig. \ref{Fig:3}, the refractive index of the region around the origin reaches the value of free space (air), making the design appealing for interfacing light and matter \cite{chang_colloquium_2018}.

On the other hand, artificial systems where the medium's index varies in a discrete fashion (unlike bulk materials), such as micro-resonators \cite{poli_selective_2015,malzard_topologically_2015}, waveguides \cite{tzortzakakis_shape-preserving_2020,weimann_topologically_2017, zeuner_observation_2015} and loudspeaker arrays \cite{rivet_constant-pressure_2018}, or time-multiplexed photonic crystals \cite{regensburger_photon_2011}, have in recent years been employed for proof-of-principle demonstrations of a plethora of wave physical phenomena. We will thus consider, as our second example, the case of such a discrete system consisting of a chain of microresonators. 

More specifically, in the stationary regime, the system is conveniently described under the coupled-mode approximation~\cite{poli_selective_2015,malzard_topologically_2015}by the following equations:
\begin{equation}\label{eq:dicrete_eq}
\begin{array}{c}
 (\gamma_n-\omega)\phi_n +\kappa(\phi_{n-1}+\phi_{n+1})=0 ,\\[0.2cm]
(\Gamma_n-\omega)E_n +\kappa(E_{n-1}+E_{n+1})=0.
\end{array}
\end{equation}
where $\omega$ is the input frequency, $\kappa$ is the nearest resonator coupling strength, $\phi_n$ is the mode amplitude in the $n$-th resonator for the distribution of resonance frequencies $\{\gamma_n\}$, while $E_n$ is the corresponding quantity for the distribution $\{\Gamma_n\}$. As before, our goal here is to create a modified medium, described by a frequency distribution $\{\Gamma_n\}\in\mathbb{R}$, supporting a solution $E_n$ which is related to the reference distribution $\{\phi_n\}$ as $E_n=A_n\exp(i\sum_{i=1}^nw_i)\phi_n$. Moreover, $\{A_n\}\in\mathbb{R}$ is the amplitude distribution (analogous to the design function), and $\{w_n\}\in\mathbb{R}$ is a distribution of coefficients, where the relative phase at the $n$-th resonator is given by $\theta_n=\sum_{i=1}^nw_n$. 

The two resonance frequency distributions are for $\{\gamma_n\},\{\Gamma_n\}\in\mathbb{R}$ related as:
 \begin{equation}\label{eq:Gamma_re}
 \begin{array}{c}
\Gamma_n=\gamma_n+\kappa\left[\eta_{n+1}^R+\xi_{n-1}^R+\frac{A_{n+1}}{A_n}(\eta_{n+1}^I\sin w_{n+1}\right.\\[0.2cm]
\left.-\eta_{n+1}^R\cos w_{n+1})- \frac{A_{n-1}}{A_n}\left(\xi_{n-1}^I\sin w_{n}+\xi_{n-1}^R\cos w_{n}\right)\right]
\end{array}
\end{equation}
where  $\phi_{n+ 1}/\phi_n=\eta_{n+ 1}^R+i\eta_{n+ 1}^I$, $\phi_{n- 1}/\phi_n=\xi_{n- 1}^R+i\xi_{n- 1}^I$, and the $w_n$'s satisfy the equation:
 \begin{equation}\label{eq:ws}
 \begin{array}{c}
A_n\left(\eta_{n+1}^I+\xi_{n-1}^I\right)-A_{n+1}(\eta_{n+1}^I\cos w_{n+1}+\eta_{n+1}^R\sin w_{n+1})\\[0.2cm]
+A_{n-1}\left(\xi_{n-1}^R\sin w_{n}-\xi_{n-1}^I\cos w_{n}\right)=0.
\end{array}
\end{equation}
We solve the equations (\ref{eq:dicrete_eq}) under boundary conditions $\phi_N=e^{ik_0n_0N\Delta x},\:\phi_{N-1}=e^{ik_0n_0(N-1)\Delta x}$, where $n_0=\frac{1}{k_0\Delta x}\sqrt{\frac{\gamma_N}{\kappa}}$, i.e., the situation is analogous to an incident propagating wave traveling to the right. The calculated parameters for the $\phi_n$ solution are substituted into Eq.\ (\ref{eq:ws}), which is then solved under the boundary conditions of $w_1=w_N=0$, in order to get the distribution $\{\Gamma_n\}$.

Figure \ref{Fig:7} shows an implementation of this idea in a chain of $N=130$ micro-resonators with a disordered distribution of resonance frequencies. Even for such an irregular reference system, Eq.\ (\ref{eq:Gamma_re}) produces an intensity confinement with a desired amplitude and width at the middle of the system, and the same intensity distribution as in the reference medium away from the center. These results provide strong indications that our theoretical study can be realized with current experimental setups.

\section{Conclusion}\label{sec:conclusion}
We have proposed a theoretical methodology for the local tailoring of light confinement in 1D inhomogeneous Hermitian dielectric media. Our approach allows us to locally design the intensity distribution of electromagnetic fields even inside disordered media, where the dependence of the modal structure on the local changes of the refractive index is highly complex. In particular, we have demonstrated that a reference and a design medium have the same unidirectional reflection and transmission coefficients. Moreover, our theory can be extended to discrete scattering systems, like for example a 1D stack of dielectric slabs, and a discrete chain of coupled microresonators. This makes our approach relevant to possible experimental realizations in the context of photonics. 

In future work, we expect to extend our methodology to two-dimenional (2D) media, in order to engineer long-lived localized states. A promising direction here would be to connect our results to commonly employed nanofabrication processes in dielectric materials, where numerical optimization techniques were typically used for optical engineering and design \cite{molesky_inverse_2018,jensen_topology_2011}. Applying our results to topological photonic media \cite{weimann_topologically_2017,ozawa_topological_2019} and media supporting bound states in the continuum \cite{marinica_bound_2008,hsu_bound_2016} presents another exciting prospect for future research. Moreover, as our approach is not limited to Hermitian dielectrics, it will be interesting to study its effectiveness in meta-materials \cite{yu_flat_2014} and in non-Hermitian media \cite{el-ganainy_non-hermitian_2018}. 

\section{Acknowledgments}
We thank Andre Brandst\" otter for sharing the transfer matrix method code for solving the 1D Helmholtz equation. This work was funded by the European Commission grant MSCA-RISE 691209, the Austrian Science Fund (FWF) grant P32300 and the FWF Lise Meitner Postdoctoral Fellowship M3011.


\clearpage
\onecolumngrid

\renewcommand{\thefigure}{S\arabic{figure}}
\renewcommand{\theequation}{S\arabic{equation}}

\setcounter{equation}{0}
\setcounter{figure}{0}
\section{Supporting information: Light confinement by local index tailoring in inhomogeneous dielectrics}

\section{I. Mapping between inhomogeneous media}\label{intro}
Let $\phi(x)$ and $E(x)$ be two solutions to the Helmholtz equation, for the same wavenumber $k_0$, and with dielectric functions $\varepsilon_\phi(x)$ and $\varepsilon_E(x)$, respectively:
\begin{equation}
\left[\frac{d^2}{dx^2}+k_0^2\varepsilon_\phi(x)\right]\phi(x)=0, \;\;
\left[\frac{d^2}{dx^2}+k_0^2\varepsilon_E(x)\right]E(x)=0.
\end{equation}
We want to map one solution onto the other, such that the two are related via a complex function $R(x)$ as:
\begin{equation}\label{eq:E_phi}
E(x) = R(x)\phi(x).
\end{equation}
By substituting this relation into the above two Helmholtz equations, we get for the dielectric function $\varepsilon_E(x)$:
\begin{equation}
\varepsilon_E(x) = \varepsilon_\phi(x)-\frac{1}{k_0^2}\left[\frac{1}{R(x)}\frac{d^2R(x)}{dx^2} +\frac{2}{\phi(x)R(x)}\frac{d\phi(x)}{dx}\frac{dR(x)}{dx} \right].
\end{equation}
The  above relation connects the two media $\varepsilon_\phi(x)$ and $\varepsilon_E(x)$. We can further express the function $R(x)$ as:
\begin{equation}
R(x) = A(x)e^{i\theta(x)},
\end{equation}
which lead us to 
\begin{align*}
\varepsilon_E(x) = \varepsilon_\phi(x)+\frac{1}{k_0^2}\left\{\left[\frac{d\theta(x)}{dx}\right]^2+2\eta_I(x)\frac{d\theta(x)}{dx} -\frac{1}{A(x)}\frac{d^2 A(x)}{dx^2} -\frac{2\eta_R(x)}{A(x)}\frac{dA(x)}{dx} \right\} \\
+i\left\{\frac{d^2\theta(x)}{dx^2}+2\left[\frac{1}{A(x)}\frac{dA(x)}{dx}+\eta_R(x)\right]\frac{d\theta(x)}{dx}+\frac{2\eta_I(x)}{A(x)}\frac{dA(x)}{dx}\right\},
\end{align*}
where $\frac{d}{dx}\ln\phi(x)=\eta_R(x)+i\eta_I(x)$. We demand that both optical media are Hermitian, which mathematically means $\mbox{Im}[\varepsilon_E(x)]=0$. This in turn means that $\theta(x)$ must satisfy the following ordinary differential equation:
\begin{align}\label{eq:theta_eq}
\frac{d^2\theta(x)}{dx^2}+2\left[\frac{1}{A(x)}\frac{dA(x)}{dx}+\eta_R(x)\right]\frac{d\theta(x)}{dx}+\frac{2\eta_I(x)}{A(x)}\frac{dA(x)}{dx}=0.
\end{align}
In the main text, we solve the above equation for $\theta(x)$ for a given $A(x)$ (where $A(x)=1$ outside of the region of interest, see below), with the boundary conditions $\frac{d\theta(x)}{dx}=0$ at the end of the simulated system, corresponding to the asymptotic regions $x\to\pm\infty$.

\section{II. Reflection and transmission coefficients of the reference and design medium}\label{sec:two}
\begin{figure}[h!]
\centering
\includegraphics[clip,width=0.75\linewidth]{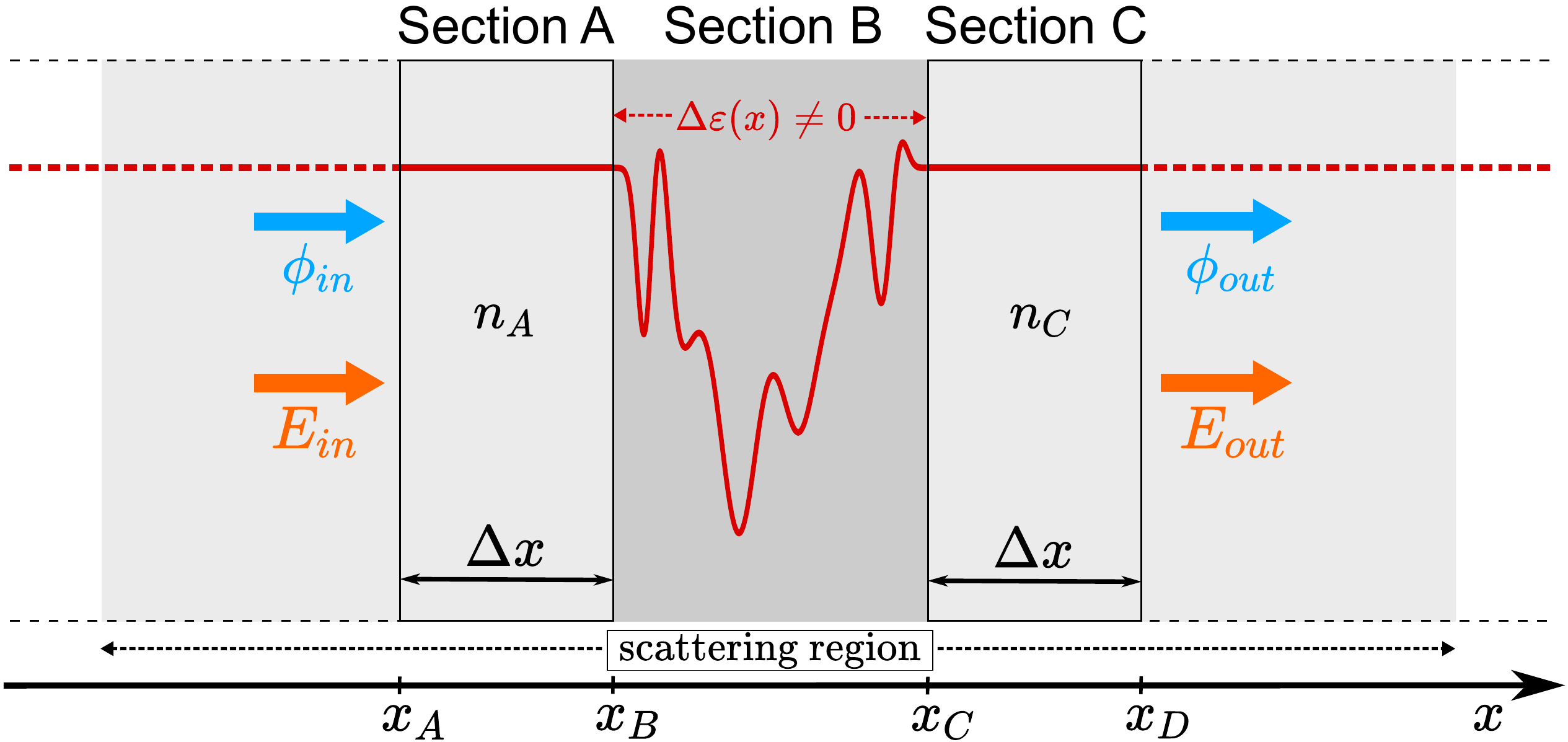}
\caption{Schematic depiction of the representation of the scattering medium that was used in deriving Eqs. (\ref{eq:r}) and (\ref{eq:t}) via the transfer matrix formalism. We focus on three sections of the reference and design medium: sections A and C with small width $\Delta x$ and effectively constant refractive indices, and section B, defined as the only part of the design medium where the difference $\Delta\varepsilon(x)$ is nonzero (shaded dark gray). The remainder of the scattering region is shaded in light gray.}\label{Fig:2A}
\end{figure} 
In the main text we consider Hermitian dielectric media, for which the function $A(x)\neq1$ only in a specific region of the scattering medium. We will now show how the difference between the dielectric functions of the reference and design media
\begin{align}\label{eq:deltaeps}
\Delta \varepsilon(x) = \frac{1}{k_0^2}\left\{\left[\frac{d\theta(x)}{dx}\right]^2+2\eta_I(x)\frac{d\theta(x)}{dx} -\frac{1}{A(x)}\frac{d^2 A(x)}{dx^2} -\frac{2\eta_R(x)}{A(x)}\frac{dA(x)}{dx} \right\}
\end{align}
vanishes in the regions where $A(x)= 1$. In these regions, the equation (\ref{eq:theta_eq}) reduces to
\begin{align}\label{eq:theta_red}
\frac{d^2\theta(x)}{dx^2}+2\eta_R(x)\frac{d\theta(x)}{dx}=0.
\end{align}
It can readily be shown that in these regions of the scattering medium the first derivative of $\theta(x)$ is given by $\frac{d\theta(x)}{dx}=C_1e^{-2\int\eta_R(x')dx'}$. For a well behaved solution $\phi(x)=\phi_Me^{i\phi_P}$ (where $\phi_M\neq 0$, and $\:\phi_P$ are real-valued functions), we can use the relation $\int\eta_R(x')dx'=C_2+\ln\phi_M(x)$ to get $\frac{d\theta(x)}{dx}=\frac{C_1e^{-2C_2}}{\phi_M^2(x)}$. As $\phi_M(x)$ does not diverge and the first derivative of $\theta(x)$ vanishes in the asymptotic regions it follows that $C_1=0$ (and/or $C_2=+\infty$), which means that $\theta(x)$ stays constant in the parts of the medium where $A(x)= 1$. This also means that for the regions where $A(x)= 1$, $\Delta\varepsilon(x)= 0$, and thus $\varepsilon_E(x)=\varepsilon_\phi(x)$. We note that by the design priciple based on Eq. (\ref{eq:E_phi}), the solutions in the reference and design media have equal moduli, i.e. $|E(x)|=|\phi(x)|$, in the regions where $A(x)=1$.

We will now use these results to relate the reflection and transmission coefficients for a right propagating incident wave of wavenumber $k_0$. In the transfer matrix method of solving the Helmholtz equation, the entire scattering region is subdivided into a large but finite number of sections of small width $\Delta x$, inside of which the refractive index value is approximated as constant (see e.g. Ref.~\cite{jirauschek} for a description of the framework). Instead of examining the propagation inside the whole scattering region, which we have done by numerically solving the equations in the main text, we now focus only on three subsections of the scattering media $n_\phi(x)$ and $n_E(x)$ (see Fig. \ref{Fig:2A}): starting at $x=x_A$ (section A), $x=x_B$ (section B) and $x=x_C$ (section C). Sections A and C have small widths $\Delta x$ and refractive indices which can be approximated as constants, given by $n_A=\sqrt{\varepsilon(x_A)}$ and $n_C=\sqrt{\varepsilon(x_C)}$, respectively, where we have used $\varepsilon(x_{A,C})=\varepsilon_\phi(x_{A,C})=\varepsilon_E(x_{A,C})$. Contrary to sections A and C, section B has a length determined by the function $A(x)$ (see main text), and is defined as the only region of the design medium where $\Delta\varepsilon(x)\neq 0$. We now examine the reflection and transmission coefficients of a wave incident on section B from the left. By using the transfer matrix method and the fact that outside of section B, $n_\phi(x)=n_E(x)$, the relationship of the scattering coefficients of the reference and the design medium can then easily be demonstrated.

In the sections A and C the electric field for the reference medium can be written as:
\begin{equation}
\begin{array}{c}
    \phi(x) = \begin{cases}
              \phi_{in}t_\phi^A e^{ik_0n_A(x-x_B)} +\phi_{in}t_\phi^A r_\phi^B e^{-ik_0n_A(x-x_B)}, & x_A< x< x_B,\\
               \phi_{in}t_\phi^A t_\phi^B e^{ik_0n_C(x-x_D)} +\phi_{in}t_\phi^A t_\phi^B r_\phi^D e^{-ik_0n_C(x-x_D)}, & x_C< x< x_D,
               \end{cases}
\end{array}
\end{equation}
while for the design case the field is
\begin{equation}
\begin{array}{c}
    E(x) = \begin{cases}
             E_{in}t_E^A e^{ik_0n_A(x-x_B)} +E_{in}t_E^A r_E^B e^{-ik_0n_A(x-x_B)}, & x_A< x< x_B,\\
               E_{in}t_E^At_E^B e^{ik_0n_C(x-x_D)} +E_{in}t_E^At_E^B r_E^D e^{-ik_0n_C(x-x_D)}, & x_C< x< x_D,
               \end{cases}
\end{array}
\end{equation}
where $\phi_{in}$ and $E_{in}$ are the (complex) amplitudes of the fields entering section A, and indices $\phi$ and $E$ mark the reference and the design medium, respectively. Since we can choose incident waves to be the same for the two cases and the two scattering potentials are equal by design for $x<x_B$, this means that $E_{in}=\phi_{in}$ and $t_\phi^A=t_E^A$. By using Eq. (\ref{eq:E_phi}) and the fact that $\theta(x)=0$ and $A(x)=1$ left of section B, we get $E(x_A<x<x_B)=\phi(x_A<x<x_B)$, which leads to:
\begin{equation}\label{eq:r}
r_E^B(k_0)=r_\phi^B(k_0)\:\:(\mbox{mod $2\pi$ in the phase})
\end{equation}
for the reflection coefficients of the reference and design media. Also, as the phase $\theta(x)$ is given by a constant $\delta_t$ outside of section B (see above), we can write $E(x_C< x< x_D)=e^{i\delta_t}\phi(x_C< x< x_D)$. As $\Delta\varepsilon=0$ outside of section B, $r_\phi^D=r_E^D$. Using these arguments, one can show that:
\begin{equation}\label{eq:t}
t_E^B(k_0)=e^{i\delta_t(k_0)}t_\phi^B(k_0),
\end{equation}
where $\delta_t(k_0)$ is the relative transmission phase shift of the solution $E(x)$ with respect to $\phi(x)$ upon passing through the section B. The Eq. (6) of the main text then follows, by using the transfer matrix formalism, from the fact that the refractive indices of the regions of the medium other than the section B are equal for the reference and design case.

\section{III. Design of arbitrarily smooth modified regions}\label{sec:three}
To demonstrate the design of arbitrarily smooth dielectric profiles in the modified region, we apply the Helmholtz equation to rewrite the dielectric function of the design medium as:
\begin{equation}\label{eq:smootheps}
\varepsilon_E(x)=-\frac{1}{k_0^2}\frac{1}{E(x)}\frac{d^2E(x)}{dx^2}.
\end{equation}
The electric field $E(x)$ can be written in polar form as $E(x)=E_M(x)e^{iE_P(x)}$, where real-valued functions $E_M(x),\:E_P(x)$ are $E_M(x)=A(x)\phi_M(x)$ and $E_P(x)=\theta(x)+\phi_P(x)$, with $\phi(x)=\phi_M(x)e^{i\phi_P}$. The Eq. (\ref{eq:smootheps}) gets then the form
\begin{equation}\label{eq:smootheps2}
\varepsilon_E(x)=-\frac{1}{k_0^2}\left[\frac{1}{E_M(x)}\frac{d^2E_M(x)}{dx^2}-\left(\frac{dE_P(x)}{dx}\right)^2+i\left(\frac{d^2E_P(x)}{dx^2}+\frac{2}{E_M(x)}\frac{dE_M(x)}{dx}\frac{dE_P(x)}{dx}\right)\right].
\end{equation}
As we require $\varepsilon_E(x)$ to be real-valued, its imaginary part must be zero, therefore:
\begin{equation}\label{eq:smootheps3}
\frac{d^2E_P(x)}{dx^2}+\frac{2}{E_M(x)}\frac{dE_M(x)}{dx}\frac{dE_P(x)}{dx}=0,
\end{equation}
and as a result
\begin{equation}\label{eq:smootheps4}
\varepsilon_E(x)=-\frac{1}{k_0^2}\left[\frac{1}{E_M(x)}\frac{d^2E_M(x)}{dx^2}-\left(\frac{dE_P(x)}{dx}\right)^2\right].
\end{equation}
If one now chooses $A(x)=\frac{p(x)}{\phi_M(x)}$, the first derivative of $E_P(x)$ can be calculated from (\ref{eq:smootheps3}) as $\frac{dE_P(x)}{dx}=\frac{C}{p^2(x)}$ (where $C$ is an integration constant, that can, for a right-propagating incident wave, be determined analytically as $C=k_0n_0\phi_M^2(+\infty)$, where we have used the fact that $E_M(+\infty)=\phi_M(+\infty)$). From the Eq. (\ref{eq:smootheps4}) we then have:
\begin{equation}\label{eq:smootheps5}
\varepsilon_E(x)=-\frac{1}{k_0^2p(x)}\frac{d^2p(x)}{dx^2}+\frac{n_0^2\phi_M^4(+\infty)}{p^4(x)}.
\end{equation}
As was shown in the previous Section, when $p(x)=\phi_M(x)$ (which means $A(x)=1$), then $\varepsilon_E(x)=\varepsilon_\phi(x)$. In the dark gray region of Fig. \ref{Fig:2A}, where $p(x)\neq\phi_M(x)$, we can see from Eq. (\ref{eq:smootheps5}) that if $p(x)$ is a sufficiently smooth function, such that $1/p(x)\cdot d^2p(x)/dx^2$ and $1/p^4(x)$ are smooth, then $\varepsilon_E(x)$ will be a smooth function, which can then, in principle, be well approximated by discrete steps of reasonable widths, in accordance to what is technically feasible with the current manufacturing resolution. 
\begin{figure}[h!]
\centering
\includegraphics[clip,width=0.5\linewidth]{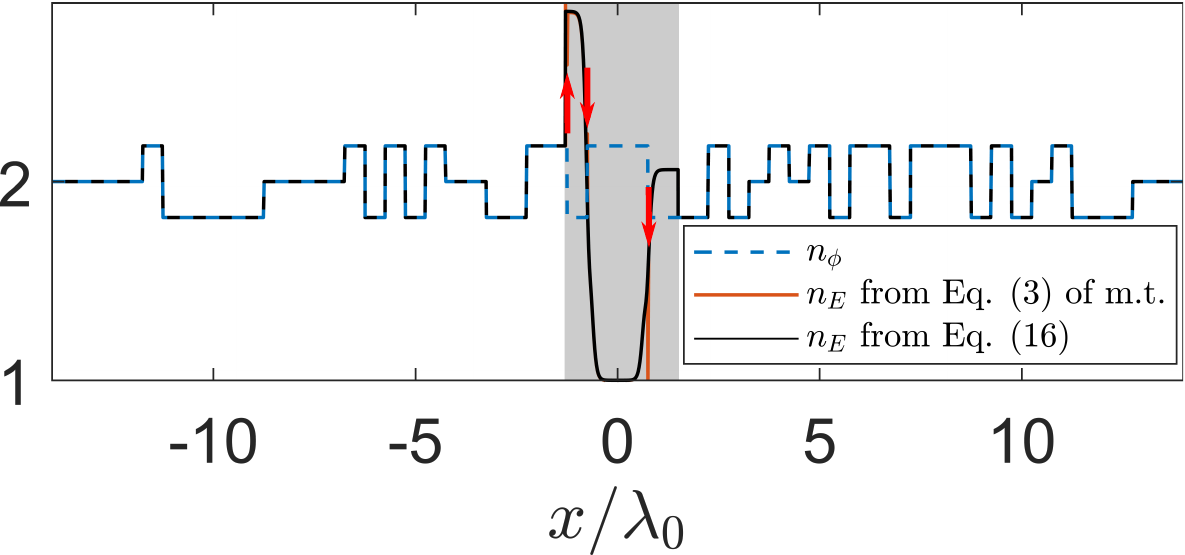}
\caption{Refractive index distributions $n_\phi(x)$ (blue, dashed) and $n_E(x)$, calculated from the Eq. (3) of the main text (orange, solid) and Eq. (\ref{eq:smootheps5}) (black, solid), for the case shown in Fig. 4 of the main text. The orange curve has very narrow ``spikes", marked by red arrows, stemming from the discontinuity of the first derivative of $A(x)$. The ``spikes" are removed when one instead uses the Eq. (\ref{eq:smootheps5}) to calculate the refractive index $n_E(x)$, as can be seen from the black curve. The gray area marks the region where $n_\phi(x)$ and $n_E(x)$ differ, i.e. $\Delta\varepsilon_E(x)\neq 0$.}\label{Fig:3A}
\end{figure} 

At the borders between the dark and light gray regions of Fig. \ref{Fig:2A}, and for a generic $A(x)$, spikes will appear, resulting in a highly oscillatory behavior of $\varepsilon_E(x)$. This is because, as the $p(x)$ approaches the function $\phi_M(x)$, the rapid oscillations in $\phi_M(x)$ will translate into rapid oscillations of $\varepsilon_E(x)$, via Eq. (\ref{eq:smootheps5}). For the Eq. (9) of the main text, this is prevented by selecting the borders of the dark and light gray regions of Fig. \ref{Fig:2A} at the position of a local minimum of $\phi_M(x)$, which means that the first term of Eq. (\ref{eq:smootheps5}) will be a constant on the left (right) side of the left (right) border, in the vicinity of $x_L$ ($x_R$). On the other hand, by selecting $p(x)=\phi_M(x_L)$ ($p(x)=\phi_M(x_R)$) inside the dark gray area near the border, the first term will vanish on the right (left) side of the left (right) border, and the second term will be matched on both sides near each of the two borders. The discrete jumps appearing in $\varepsilon_E(x)$ (and consequently in $n_E(x)$) at the borders, which are compatible or even desirable from a fabrication perspective, will then be a result of the first term of Eq. (\ref{eq:smootheps5}) (a finite-valued constant on the left (right), and zero on the right (left) side of the left (right) border) changing its value when crossing the borders. As a result the fast oscillations or spikes will be avoided.

Equation (\ref{eq:smootheps5}) is an analytical formula of the expression (3) of the main text, and is also equivalent to the Helmholtz equation for the field amplitude in the design medium. As such, it has the advantage that it requires us to numerically solve only the Helmholtz equation for $\phi(x)$, and not the ordinary differential equation for the phase $\theta(x)$ (\ref{eq:theta_eq}), in order to calculate the refractive index $n_E(x)$. A disadvantage is that the relationship between $n_\phi(x)$ and $n_E(x)$ is not obvious based on such formulation. 

Another relevant point to our example of a discrete dielectric stack is that the $n_E(x)$ calculated from the expression (3) of the main text has three ``spikes" stemming from the discontinuities of the first derivative of $A(x)$ at the points where $n_\phi(x)$ is discontinuous. This can be seen in the orange line of Fig. \ref{Fig:3A}. The ``spikes" disappear when using Eq. (\ref{eq:smootheps5}) instead, as can be seen in the black line of Fig. \ref{Fig:3A}.
\bibliography{references}
\clearpage
\newpage



\end{document}